\documentclass{article}
\usepackage{spconf,amsmath,graphicx}
\usepackage{times,epsfig,enumitem,amssymb,subcaption}

\usepackage{enumitem}
\setlist{nosep, leftmargin=14pt}

\usepackage{mwe} 

\title{Sharp-GAN: SHARPNESS LOSS REGULARIZED GAN FOR HISTOPATHOLOGY IMAGE SYNTHESIS}
\name{Sujata Butte$^\ast$, Haotian Wang$^\ast$\thanks{$^\ast$ First authors contributed equally.}, Min Xian$^\dagger$, Aleksandar Vakanski\thanks{$^\dagger$ Corresponding author.}}
\address{University of Idaho, Idaho, USA}
\begin{document}
%
\maketitle
\begin{abstract}
Existing deep learning-based approaches for histopathology image analysis require large annotated training sets to achieve good performance; but annotating histopathology images is slow and resource-intensive. Conditional generative adversarial networks have been applied to generate synthetic histopathology images to alleviate this issue, but current approaches fail to generate clear contours for overlapped and touching nuclei. In this study, We propose a sharpness loss regularized generative adversarial network to synthesize realistic histopathology images. The proposed network uses normalized nucleus distance map rather than the binary mask to encode nuclei contour information. The proposed sharpness loss enhances the contrast of nuclei contour pixels. The proposed method is evaluated using four image quality metrics and segmentation results on two public datasets. Both quantitative and qualitative results demonstrate that the proposed approach can generate realistic histopathology images with clear nuclei contours.

\end{abstract}


\section{Introduction}
\label{sec:intro}

Nuclei analysis is a critical step in computational pathology for quantitatively cancer grading. Most current nuclei segmentation approaches use deep learning-based end-to-end framework to map histopathology images and dense predictions \cite{naylor2018segmentation, graham2019hover, wang2020bending, kumar2017dataset, zhou2019cia}, and requires large accurately labeled training sets. However, the manual annotation of nuclei is time-consuming, labor-intensive and expensive. Only experienced pathologists who have been trained for years can annotate the image accurately; and, on an average, it costs about two hours for an expert to annotate an image patch with 600 nuclei \cite{hou2019robust}. 

To overcome the challenge, Zhou \textit{et al.} \cite{zhou2017evaluation} proposed to re-distribute segmented nuclei to generate more nuclei masks and histopathology images. Those masks may not be the true masks of the original nuclei, and no new nuclei shapes were generated. Hou \textit{et al.} \cite{hou2019robust} proposed a GAN network that learned the texture and shape information from the real nuclei images, and synthesized realistic histopathology images from binary image masks. However, few studies proposed an image-to-image translation network that converts the nuclei labels to images in histopathology analysis. 

\begin{figure}[t]

\begin{minipage}[b]{0.15\linewidth}
  \centering
  \centerline{\includegraphics[width=1.4cm, height=1.2cm]{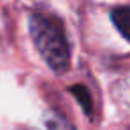}}
\end{minipage}
\begin{minipage}[b]{0.15\linewidth}
  \centering
  \centerline{\includegraphics[width=1.4cm, height=1.2cm]{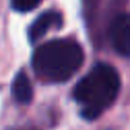}}
\end{minipage}
\begin{minipage}[b]{0.15\linewidth}
  \centering
  \centerline{\includegraphics[width=1.4cm, height=1.2cm]{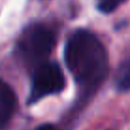}}
\end{minipage}
\begin{minipage}[b]{0.15\linewidth}
  \centering
  \centerline{\includegraphics[width=1.4cm, height=1.2cm]{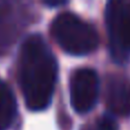}}
\end{minipage}
\begin{minipage}[b]{0.15\linewidth}
  \centering
  \centerline{\includegraphics[width=1.4cm, height=1.2cm]{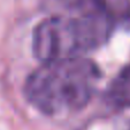}}
\end{minipage}
\begin{minipage}[b]{0.15\linewidth}
  \centering
  \centerline{\includegraphics[width=1.4cm, height=1.2cm]{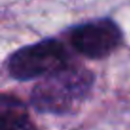}}
\end{minipage}
\hfill
\begin{minipage}[b]{0.15\linewidth}
  \centering
  \centerline{\includegraphics[width=1.4cm, height=1.2cm]{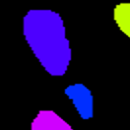}}
\end{minipage}
\hfill
\begin{minipage}[b]{0.15\linewidth}
  \centering
  \centerline{\includegraphics[width=1.4cm, height=1.2cm]{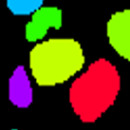}}
\end{minipage}
\hfill
\begin{minipage}[b]{0.15\linewidth}
  \centering
  \centerline{\includegraphics[width=1.4cm, height=1.2cm]{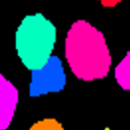}}
\end{minipage}
\hfill
\begin{minipage}[b]{0.15\linewidth}
  \centering
  \centerline{\includegraphics[width=1.4cm, height=1.2cm]{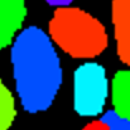}}
\end{minipage}
\hfill
\begin{minipage}[b]{0.15\linewidth}
  \centering
  \centerline{\includegraphics[width=1.4cm, height=1.2cm]{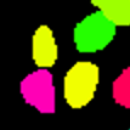}}
\end{minipage}
\hfill
\begin{minipage}[b]{0.15\linewidth}
  \centering
  \centerline{\includegraphics[width=1.4cm, height=1.2cm]{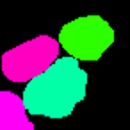}}
\end{minipage}
\hfill
\begin{minipage}[b]{0.15\linewidth}
  \centering
  \centerline{\includegraphics[width=1.4cm, height=1.2cm]{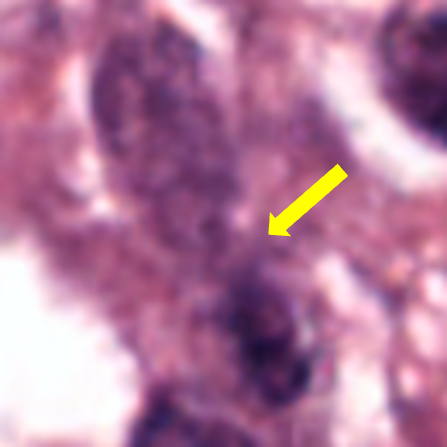}}
\end{minipage}
\hfill
\begin{minipage}[b]{0.15\linewidth}
  \centering
  \centerline{\includegraphics[width=1.4cm, height=1.2cm]{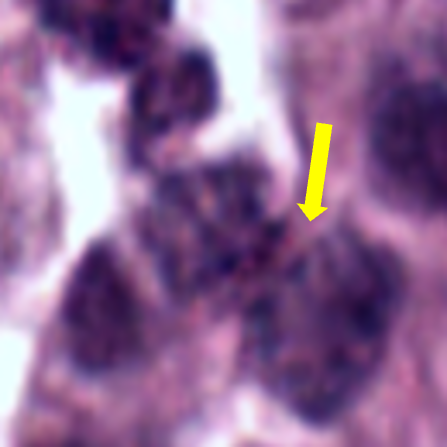}}
\end{minipage}
\hfill
\begin{minipage}[b]{0.15\linewidth}
  \centering
  \centerline{\includegraphics[width=1.4cm, height=1.2cm]{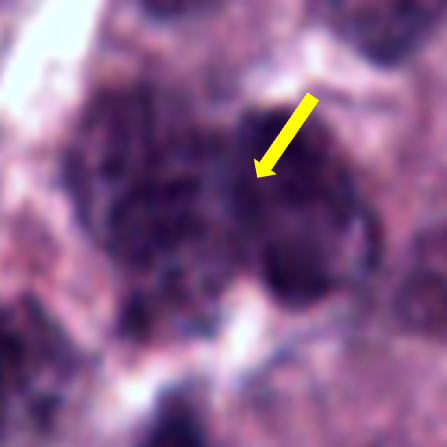}}
\end{minipage}
\hfill
\begin{minipage}[b]{0.15\linewidth}
  \centering
  \centerline{\includegraphics[width=1.4cm, height=1.2cm]{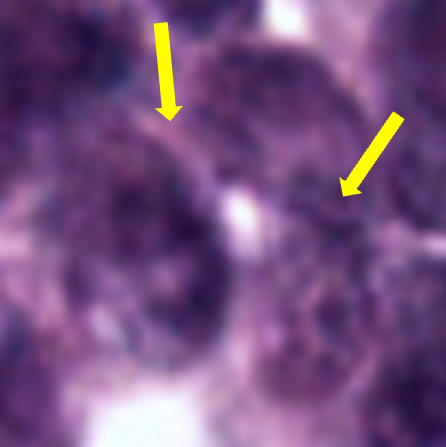}}
\end{minipage}
\hfill
\begin{minipage}[b]{0.15\linewidth}
  \centering
  \centerline{\includegraphics[width=1.4cm, height=1.2cm]{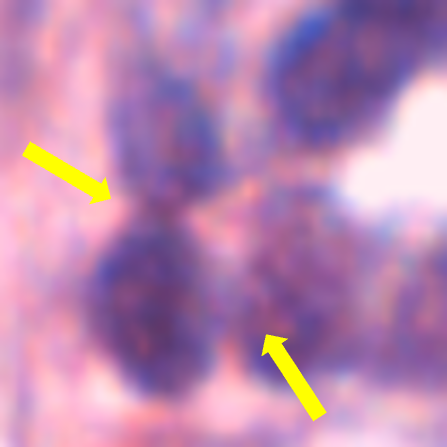}}
\end{minipage}
\hfill
\begin{minipage}[b]{0.15\linewidth}
  \centering
  \centerline{\includegraphics[width=1.4cm, height=1.2cm]{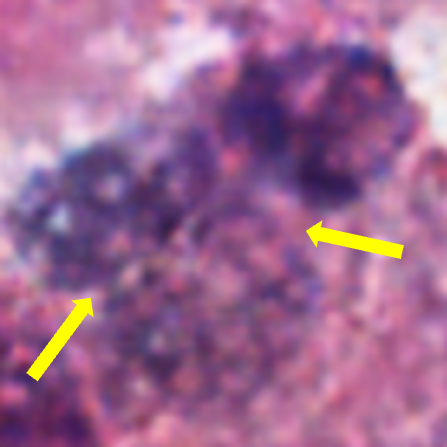}}
\end{minipage}
\hfill
\begin{minipage}[b]{0.15\linewidth}
  \centering
  \centerline{\includegraphics[width=1.4cm, height=1.2cm]{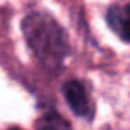}}
\end{minipage}
\hfill
\begin{minipage}[b]{0.15\linewidth}
  \centering
  \centerline{\includegraphics[width=1.4cm, height=1.2cm]{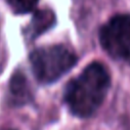}}
\end{minipage}
\hfill
\begin{minipage}[b]{0.15\linewidth}
  \centering
  \centerline{\includegraphics[width=1.4cm, height=1.2cm]{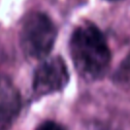}}
\end{minipage}
\hfill
\begin{minipage}[b]{0.15\linewidth}
  \centering
  \centerline{\includegraphics[width=1.4cm, height=1.2cm]{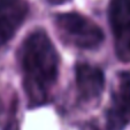}}
\end{minipage}
\hfill
\begin{minipage}[b]{0.15\linewidth}
  \centering
  \centerline{\includegraphics[width=1.4cm, height=1.2cm]{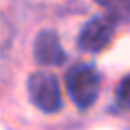}}
\end{minipage}
\hfill
\begin{minipage}[b]{0.15\linewidth}
  \centering
  \centerline{\includegraphics[width=1.4cm, height=1.2cm]{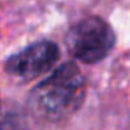}}
\end{minipage}

\vspace{-0.3em}
\caption{First row: Real image patches. Second row: nuclei masks, different colors represent different nuclei. Third row: synthesized image patches from pix2pix-GAN \cite{isola2017image}. Forth row: our synthesized image patches}
\label{fig:res}
\vspace{-0.3em}
\end{figure}

\begin{figure*}[]
\small
\begin{center}
   \includegraphics[width=0.95\linewidth]{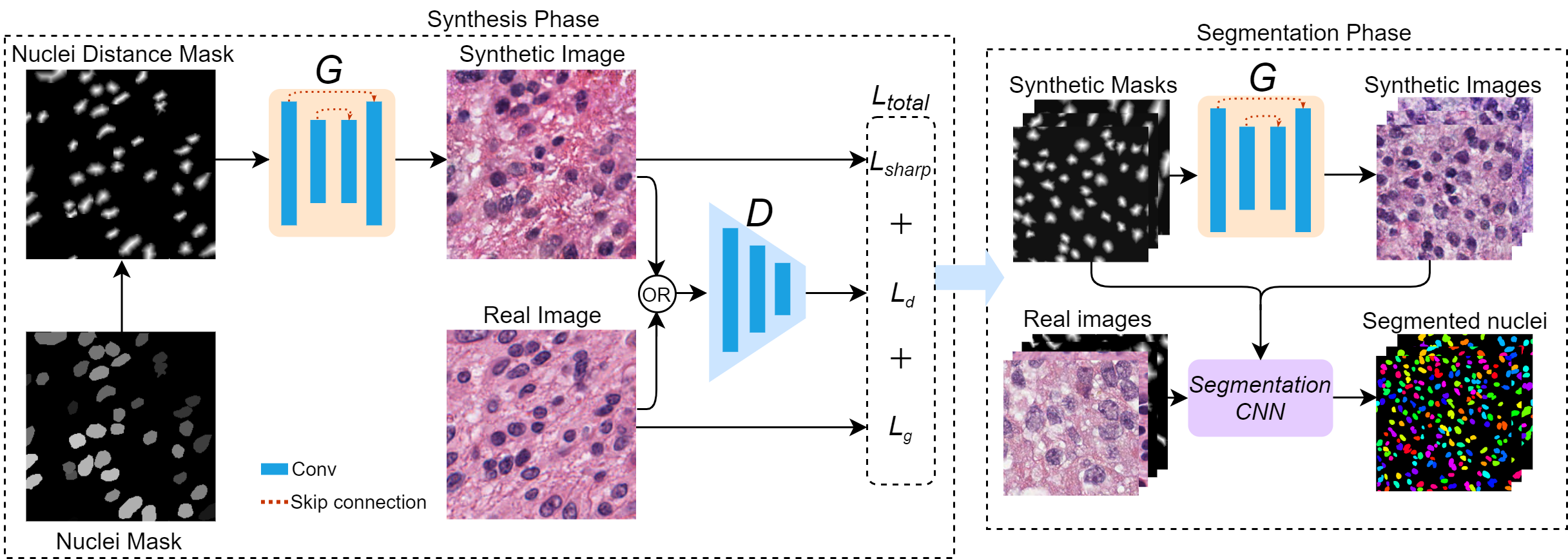}
\end{center}
\vspace{-0.3em}
   \caption{Overview of the proposed Sharp-GAN.}
\label{fig:long}
\label{fig:onecol}
\vspace{-0.3em}
\end{figure*}

Conditional image synthesis has been studied widely in many fields, e.g, class-conditional models learn to synthesis images using category labels \cite{odena2017conditional}, image-to-image translation \cite{isola2017image, zhu2017unpaired}. Comparing to traditional image synthesis methods \cite{ hertzmann2001image}, deep Learning-based methods generated more realistic images. However, existing deep approaches still have limits in synthesizing histopathology images with clear nuclei boundaries. As shown Fig. 1, the approach in \cite{isola2017image} cannot generate clear boundaries between closely clustered nuclei.  

To solve this issue, we proposed a sharpness loss regularized GAN, namely Sharp-GAN to generate realistic histopathology images with clear nuclei contours. The proposed Sharp-GAN synthesizes histopathology images using distance maps of nuclei which provides full contours of nuclei. The conventional binary masks cannot accurately define the shared contours of touching nuclei. We propose the sharpness loss to reveal the contours between touching nuclei by enhancing the contrast of the contour pixels. The sharpness loss defines large penalty for contours points with small contrast.

\section{Proposed Method}
\label{sec:methods}
The generator $\mathcal{G}$ in the proposed network learns a mapping from observed nuclei distance map $x \in \Bbb R_{\ge 0}^{mn} $ to realistic histopathology image $y$ that can 'fool' the discriminator $\mathcal{D}$, where $m$ and $n$ is the width and height of the image, respectively. $x_{i,j}$ in the distance map is defined as the Euclidean distance from $(i,j)$ to a nucleus's centroid. The discriminator $\mathcal{D}$ is trained to detect 'fake' images produced by $\mathcal{G}$. The proposed network is shown in Fig. 2.
\subsection{Loss function}
Conditional GANs \cite{isola2017image} tends to produce blur nucleus boundaries, and have difficulties in distinguish closely clustered nuclei. We proposed the sharpness loss to enhance the contrast of nuclei contour pixels, and the total loss is defined by
\begin{equation}
\mathcal{L(G, D)} = \mathbb{E}_{x,y}[L_1] + \mathbb{E}_{x}[L_2 + \beta L_{sharp}]
\end{equation}
where
\begin{equation}
L_{1} = log\mathcal{D}(x,y),
\end{equation}
\begin{equation}
L_2 = log(1-\mathcal{D}(x,\mathcal{G}(x)))
\end{equation}

The goal is to obtain an optimal generator $\mathcal{G^*}$ that minimizes the total loss $\mathcal{L}$ against an optimal $\mathcal{D^*}$ that maximizes the total loss.  $L_{sharp}$ denoted the proposed sharpness loss, the parameter $\beta$ controls contribution of the sharpness loss. The generator network follows the U-Net-based pixel2pixel GAN \cite{isola2017image}. Inspired by Graph cuts approach\cite{937505}, we proposed the sharpness loss to enhance the contrast of contour pixels of nuclei (rows 3\&4 in Fig. 1). The contour pixels are obtained by multiplying a binary contour map with the synthetic image. The binary contour map is generated by using the binary nuclei masks. Let $c$ be a contour map, and $\hat{y}$ be an synthesized image of $\mathcal{G}(x)$ and the sharpness loss is defined by
\begin{equation}
L_{sharp}(c,\hat{y})= \frac{1}{mn} \sum_{i,j, c_{i,j}\ne 0}S(i,j)
\end{equation}

\hspace{-2em} where $S(i, j)$ is the discrete sharpness at position $(i,j)$, and is given by
\begin{equation}
S(i,j) = \sum_{(p,q) \in {N_{i,j}} } exp (\frac{-(g_{i,j}-g_{p,q})^2}{2\lambda^2}) \cdot \frac{1}{dist}
\end{equation}

In Eq. (5), $g$ is a gray scale image transformed from $\hat{y}$. $g_{i,j}$ and $g_{p,q}$ are the intensities of pixels at position (i, j) and (p,q), respectively. $N_{i,j}$ is the set of the neighboring positions of $(i,j)$. $\lambda$ is a threshold to determine the degrees of contrast. Large $\lambda$ leads to contours with higher contrast. \textit{dist} is the Manhattan Distance between two contour pixels. The proposed loss gives high penalty for continues pixels when $\left|g_{i,j}-g_{p,q}\right| \textless \lambda$; and it defines small penalty when  $\left|g_{i,j}-g_{p,q}\right| \textgreater\lambda$.  We apply the eight nearest neighboring system to define neighbors. 

\begin{figure}

\begin{minipage}[b]{0.24\linewidth}
  \centering
  \centerline{\includegraphics[width=\linewidth]{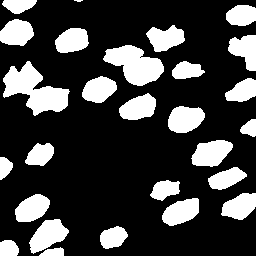}}
\end{minipage}
\hfill
\begin{minipage}[b]{0.24\linewidth}
  \centering
  \centerline{\includegraphics[width=\linewidth]{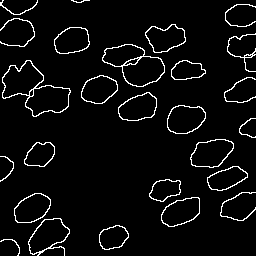}}
\end{minipage}
\hfill
\begin{minipage}[b]{0.24\linewidth}
  \centering
  \centerline{\includegraphics[width=\linewidth]{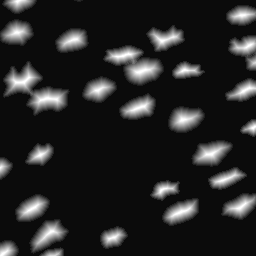}}
\end{minipage}
\hfill
\begin{minipage}[b]{0.24\linewidth}
  \centering
  \centerline{\includegraphics[width=\linewidth]{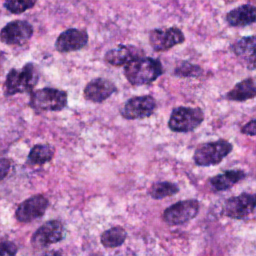}}
\end{minipage}
\hfill
\vspace{-0.3em}
\caption{A sample of (a) binary mask, (b) contour map, (c) distance map, and (d) synthetic image patch.}
\label{fig:res}
\vspace{-0.3em}
\end{figure}
\subsection{Nucleus distant map}
Many existing approaches \cite{isola2017image} used binary image masks to input conditional GAN to generate synthesis images. However, in the histopathology image, many nuclei are closely clustered, and the binary masks cannot show the boundary between two touching nuclei. If we use the binary masks to train the generator, the generator cannot know the accurate boundaries of nuclei and produce synthetic images with blur boundaries. To address this issue, we proposed to generate a distance map $x$ from binary masks to clearly separate touching nuclei. 

Let $b$ be a binary mask of an image. If $b(i,j)$ is 0, and the value of the distance map at position $(i, j): x(i,j)$ is 0; otherwise, $x(i,j)$ will be the the Euclidean distance from $(i,j)$ to the centroid of the nucleus that contains $(i,j)$. Fig. 3(d) shows a sample of nuclei distance mask.

We employed the nuclei mask generation method in \cite{hou2019robust}. The method randomly generates nucleus-like polygons with variable size and irregularities. The generated polygons can be set to overlap and touch to each other. Fig. 3(a) shows an sample synthetic image with nucleus-like polygons. 

\section{EXPERIMENTAL RESULTS}
\label{sec:EXPERIMENTS AND RESULTS}

\subsection{Datasets, metrics and setting}

\textbf{Dataset}. We evaluate the performance of the proposed approach using two public histopathology datasets, MICCAI 2015 Digital Pathology challenge (CPM-15) and MICCAI 2017 Digital Pathology challenge (CPM-17) \cite{vu2019methods}. CPM-15 contains 15 histopathology images with a total number of 2905 nuclei. CPM-17 contains 32 histopathology images with a total number of 7570 nuclei. The magnifications of these two datasets contain both 20$\times$ and 40$\times$. We combine two datasets in our experiments.

\textbf{Evaluation metrics}. Four image quality assessment metric, structural similarity index (SSIM) \cite{wang2004image}, feature similarity index (FSIM)\cite{zhang2011fsim}, gradient magnitude similarity deviation (GMSD) \cite{xue2013gradient}, normalized root mean square error (NRMSE) are used to evaluate the quality of synthesized images. 

We employed four metrics, detection quality (DQ) \cite{kirillov2019panoptic}, segmentation quality (SQ) \cite{kirillov2019panoptic}, panoptic quality (PQ) \cite{kirillov2019panoptic}, aggregated jaccard index (AJI) \cite{kumar2017dataset} to evaluate the performance of segmentation approaches trained using different datasets. These four metrics have been applied in the nuclei segmentation  \cite{graham2019hover, wang2020bending}.

\textbf{Training process}. We feed images with size of 256$\times$256 into the proposed network. We used random crop, flip, rotation for data augmentation. We set the batch size to 16, the initial learning rate to 0.0001, and epochs to 500 for model training. 

\subsection{Image quality assessment}

To validate the effectiveness of the proposed distance map, we compare the proposed network using binary mask input (GAN-binary) with the proposed network with the distance map (GAN-DIST) input. In the experiments, GAN-binary model is trained using the CPM15\&17 training set with nuclei binary masks, and GAN-DIST is trained using distance maps. Four image quality metrics, SSIM, FSIM, GMSD, and NRMSE, are employed to quantitatively evaluate the quality of synthetic images. Table 1 shows the image quality measurements of all test images. We note that GAN-DIST outperforms GAN-Binary in all metrics, i.e., the SSIM, FSIM, GMSD, NRMSE results improved 18.0\%, 3.8\%, 9.2\%, 18.1\%, respectively. The results demonstrate that using distance maps as input improves the image quality of synthetic images.

To demonstrate the effectiveness of the proposed sharpness term, we compared our model with the sharpness term (Sharp-GAN) and without the sharpness term (GAN-DIST). The two models use distance maps as input. As shown in Table 1, Sharp-GAN outperformes the GAN-DIST in all four metrics. SSIM, FSIM, GMSD, NRMSE of Sharp-GAN are 11.2\%, 3.9\%, 10.8\%, 11.1\% higher than those of the GAN-DIST, respectively.  

\begin{figure}[]
\small
\begin{minipage}[b]{0.24\linewidth}
  \centering
  \centerline{\includegraphics[width=\linewidth]{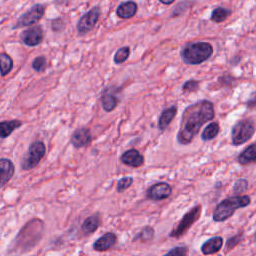}}
\end{minipage}
\hfill
\begin{minipage}[b]{0.24\linewidth}
  \centering
  \centerline{\includegraphics[width=\linewidth]{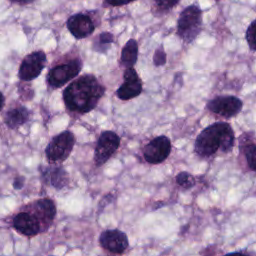}}
\end{minipage}
\hfill
\begin{minipage}[b]{0.24\linewidth}
  \centering
  \centerline{\includegraphics[width=\linewidth]{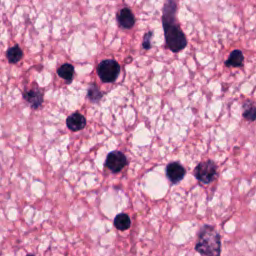}}
\end{minipage}
\hfill
\begin{minipage}[b]{0.24\linewidth}
  \centering
  \centerline{\includegraphics[width=\linewidth]{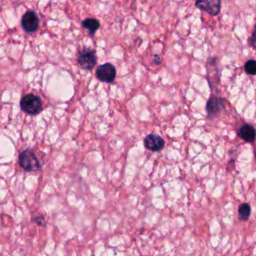}}
\end{minipage}
\hfill

\begin{minipage}[b]{0.24\linewidth}
  \centering
  \centerline{\includegraphics[width=\linewidth]{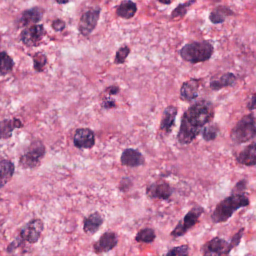}}
\end{minipage}
\hfill
\begin{minipage}[b]{0.24\linewidth}
  \centering
  \centerline{\includegraphics[width=\linewidth]{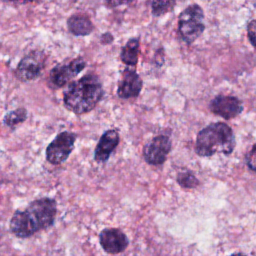}}
\end{minipage}
\hfill
\begin{minipage}[b]{0.24\linewidth}
  \centering
  \centerline{\includegraphics[width=\linewidth]{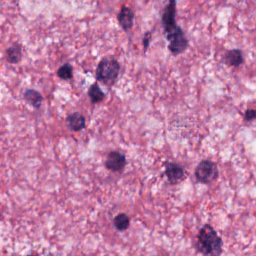}}
\end{minipage}
\hfill
\begin{minipage}[b]{0.24\linewidth}
  \centering
  \centerline{\includegraphics[width=\linewidth]{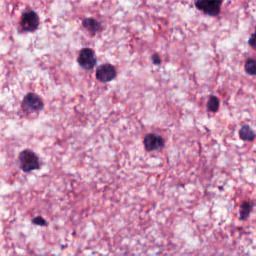}}
\end{minipage}
\hfill
\vspace{-0.3em}
   \caption{Samples of original histopathology patches (first row) with the corresponding synthetic patches generated by our methods (second row).}
\label{fig:long}
\label{fig:onecol}
\vspace{-0.3em}
\end{figure}

Fig. 4 are samples of the synthesized image patches by our model; we noted that the foreground nuclei are close to the original histopathology patches, and the background shows a realistic texture. Fig. 1 fourth row shows some examples of our synthesized patches, it shows that our model can produce clear nuclei boundary in the overlapped nuclei regions.

\begin{table}[]
\small
\caption{Image quailty assessment using CPM15\&17 dataset.}
\vspace{-2em}
\begin{center}
\begin{tabular}{|c|c|c|c|c|}
\hline
Model & SSIM$\uparrow$ & FSIM$\uparrow$ & GMSD$\downarrow$ & NRMSE$\downarrow$\\
\hline\hline
        GAN-Binary  & 0.577 & 0.735 & 0.178 & 0.306 \\
        GAN-DIST & 0.681 & 0.763 & 0.163 & 0.259 \\
        Sharp-GAN & \textbf{0.756} & \textbf{0.793} & \textbf{0.147} & \textbf{0.230} \\
\hline
\end{tabular}
\end{center}
\vspace{-0.3em}
\end{table}

\begin{figure}[t]
\small
\rotatebox{90}{\hspace{0.7cm}(a)}
\begin{minipage}[b]{0.23\linewidth}
  \centering
  \centerline{\includegraphics[width=\linewidth, height=\linewidth]{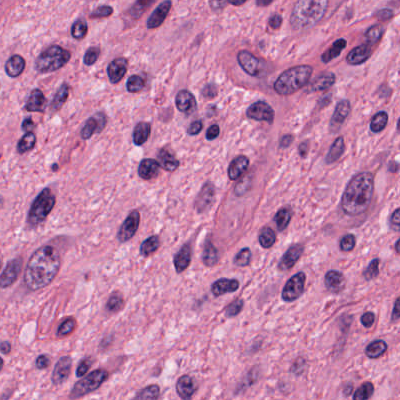}}
\end{minipage}
\hfill
\begin{minipage}[b]{0.23\linewidth}
  \centering
  \centerline{\includegraphics[width=\linewidth]{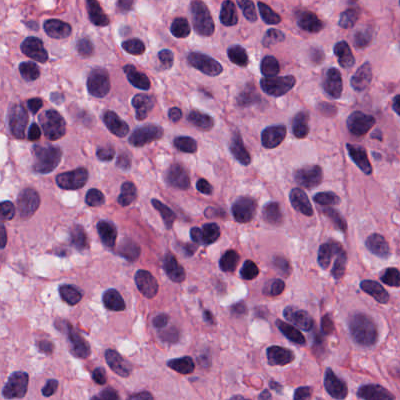}}
\end{minipage}
\hfill
\begin{minipage}[b]{0.23\linewidth}
  \centering
  \centerline{\includegraphics[width=\linewidth]{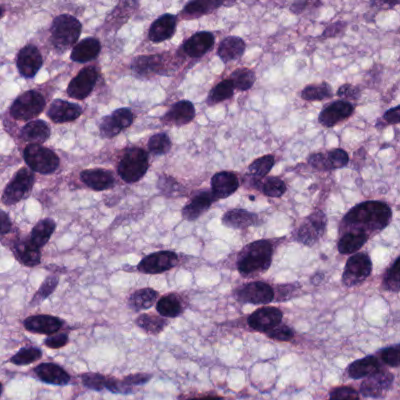}}
\end{minipage}
\hfill
\begin{minipage}[b]{0.23\linewidth}
  \centering
  \centerline{\includegraphics[width=\linewidth]{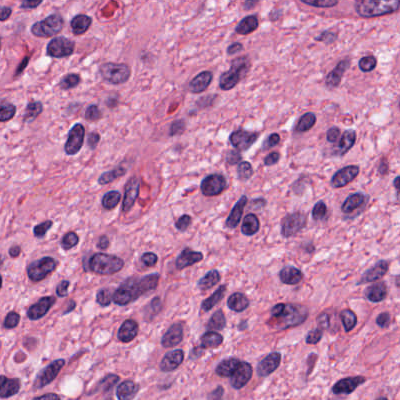}}
\end{minipage}
\hfill

\rotatebox{90}{\hspace{0.7cm}(b)}
\begin{minipage}[b]{0.23\linewidth}
  \centering
  \centerline{\includegraphics[width=\linewidth]{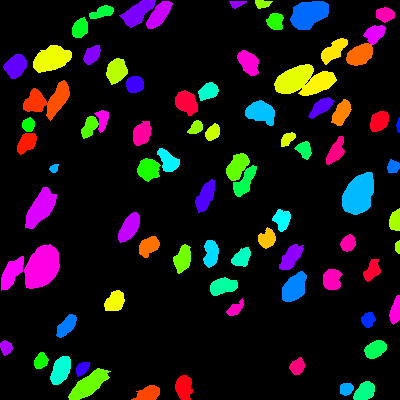}}
\end{minipage}
\hfill
\begin{minipage}[b]{0.23\linewidth}
  \centering
  \centerline{\includegraphics[width=\linewidth]{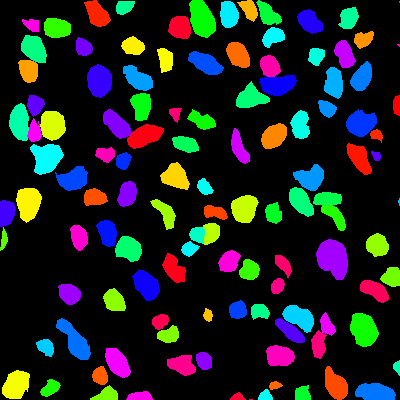}}
\end{minipage}
\hfill
\begin{minipage}[b]{0.23\linewidth}
  \centering
  \centerline{\includegraphics[width=\linewidth]{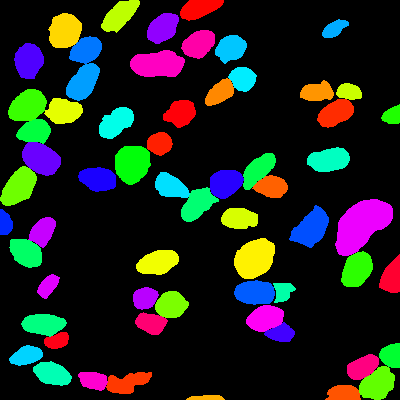}}
\end{minipage}
\hfill
\begin{minipage}[b]{0.23\linewidth}
  \centering
  \centerline{\includegraphics[width=\linewidth]{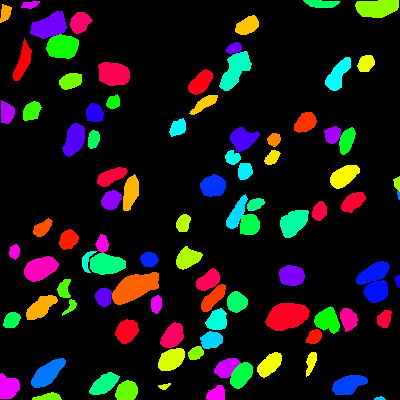}}
\end{minipage}
\hfill

\rotatebox{90}{\hspace{0.7cm}(c)}
\begin{minipage}[b]{0.23\linewidth}
  \centering
  \centerline{\includegraphics[width=\linewidth]{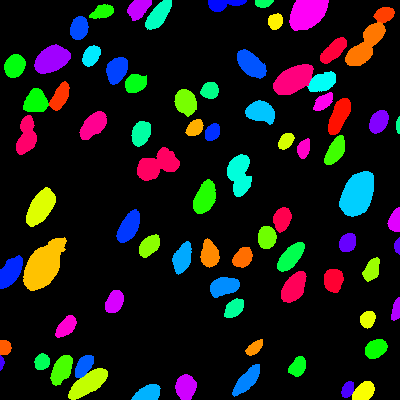}}
\end{minipage}
\hfill
\begin{minipage}[b]{0.23\linewidth}
  \centering
  \centerline{\includegraphics[width=\linewidth]{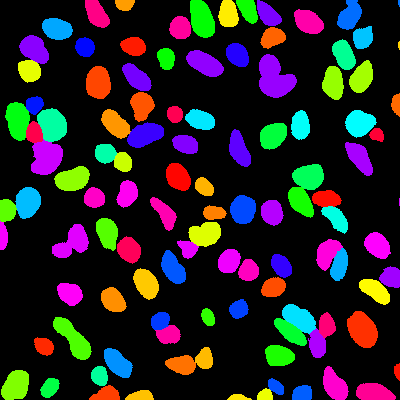}}
\end{minipage}
\hfill
\begin{minipage}[b]{0.23\linewidth}
  \centering
  \centerline{\includegraphics[width=\linewidth]{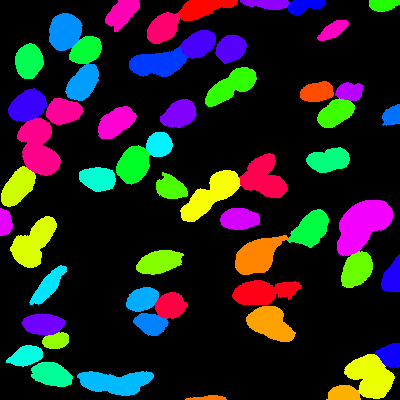}}
\end{minipage}
\hfill
\begin{minipage}[b]{0.23\linewidth}
  \centering
  \centerline{\includegraphics[width=\linewidth]{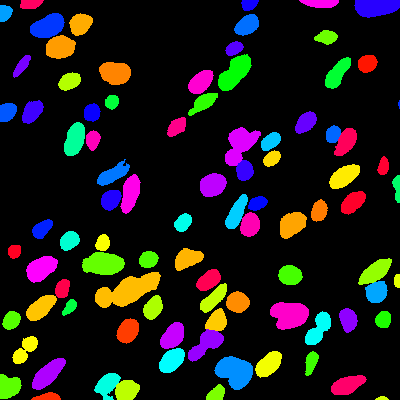}}
\end{minipage}
\hfill

\rotatebox{90}{\hspace{0.7cm}(d)}
\begin{minipage}[b]{0.23\linewidth}
  \centering
  \centerline{\includegraphics[width=\linewidth]{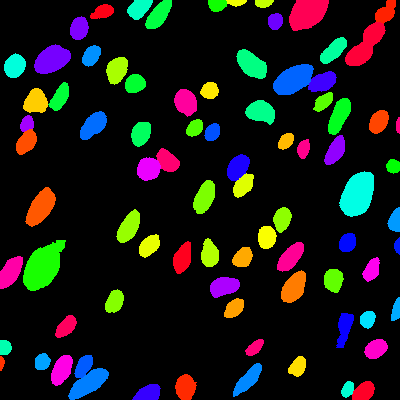}}
\end{minipage}
\hfill
\begin{minipage}[b]{0.23\linewidth}
  \centering
  \centerline{\includegraphics[width=\linewidth]{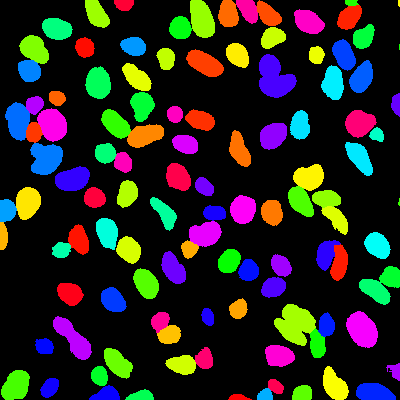}}
\end{minipage}
\hfill
\begin{minipage}[b]{0.23\linewidth}
  \centering
  \centerline{\includegraphics[width=\linewidth]{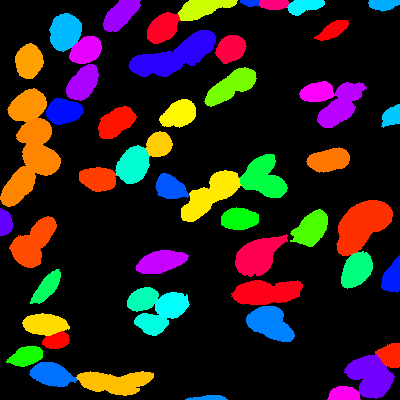}}
\end{minipage}
\hfill
\begin{minipage}[b]{0.23\linewidth}
  \centering
  \centerline{\includegraphics[width=\linewidth]{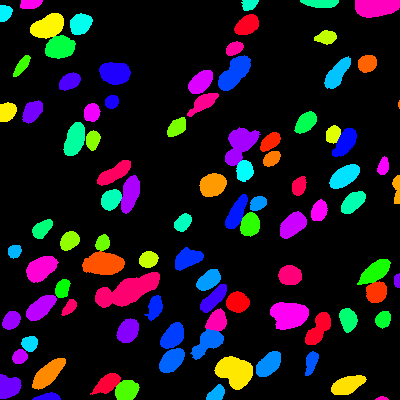}}
\end{minipage}
\hfill
\vspace{-0.3em}
   \caption{Segmentation results. (a) Original images; (b) ground truth; (c) U-Net results using synthetic and real images; and (d) U-Net results using synthetic images only. Different colors represent different nuclei.}
\label{fig:long}
\label{fig:onecol}
\vspace{-0.3em}
\end{figure}

\begin{table}[t]
\small
\caption{Segmentation results on CPM-15\&17 test set.}
\vspace{-2em}
\begin{center}
\begin{tabular}{|l|l|c|c|c|c|c|}
\hline
Model & Training & DQ & SQ & PQ & AJI \\
\hline\hline
SegNet & R  & 0.765 & 0.759 & 0.584 & 0.588\\
                & S (3k) & 0.686 & 0.757 & 0.523 & 0.510\\
                & R+S (3k) & 0.780 & 0.769 & 0.602 & 0.608 \\
                & S (10k) & 0.732 & 0.759 & 0.557 & 0.545\\
                & R+S (10k) & \textbf{0.806} & 0.\textbf{772} & \textbf{0.608} & \textbf{0.623} \\
\hline
U-Net & R & 0.825 & 0.795 & 0.658 & 0.670\\
                & S (3k) & 0.755 & 0.738 & 0.562 & 0.593\\
                & R+S (3k) & 0.837 & 0.800 & 0.671 & 0.675 \\
                & S (10k) & 0.804 & 0.788 & 0.634 & 0.616\\
                & R+S (10k) & \textbf{0.841} & \textbf{0.797} & \textbf{0.669} & \textbf{0.677} \\
\hline
\end{tabular}
\end{center}
\vspace{-2em}
\end{table}

\subsection{Application to nuclei segmentation}
To demonstrate the effectiveness of the proposed Sharp-GAN, we evaluate segmentation models trained using synthetic images generated from Sharp-GAN. Two semantic segmentation networks, U-Net \cite{ronneberger2015u} and SegNet \cite{badrinarayanan2017segnet} are employed to segment the nuclei foreground from the background. 
 
We compare the segmentation networks using different training sets. R contains only real images from the CPM15\&17 training set; S denotes as set of pure synthetic images; and R+S denotes a combined set of real and synthetic images. As shown in Table 2, the results of both U-Net and SegNet trained using 3k synthetic images are only slightly worse than the results of models trained using real data. While we train the models using a combination of synthetic data and real data, their results of the four metrics are all higher than those of the models trained using only real images. 

Meanwhile, we synthesize two datasets, one with 10k images and another with 3k images, and train U-Net and SegNet using them. Table 2 shows that the segmentation models trained using 10k synthetic images and real images outperform the models training using a combinations of 3k synthetic images and real images. Fig. 5 shows four segmentation examples of U-Net. They show that the model trained using only synthetic images is comparable to the model trained using both synthetic and real images.

\section{Conclusion}
\label{sec:Discussion and Conclusion}

In this paper, we propose the Sharp-GAN to generate realistic histopathology images. It uses nucleus distance map as input to clearly define the contours between overlapped and touching nuclei; and we propose the sharpness term to enhance the contrast of nuclei contours. Quantitative evaluation demonstrates that the proposed approach can generate histopathology images with high qualities. The segmentation results demonstrate that the models, e.g., U-Net and SegNet trained using both synthetic and real images significantly outperform the those trained using only real images.

\bibliographystyle{IEEEbib}
\bibliography{strings,refs}

\end{document}